\documentclass[reprint,amsmath,amssymb,aps]{revtex4-1}
\usepackage{graphicx,epsfig,dcolumn,bm,epic,eepic,float}
\usepackage{amsmath}
\usepackage{latexsym}
\usepackage{color}
\usepackage{cancel}
\usepackage{makeidx,shortvrb,latexsym}
\usepackage{epstopdf}
\usepackage{graphicx}
\pdfoutput=1
\begin{document}
\unitlength 1 cm
\newcommand{\nn}{\nonumber}
\newcommand{\vk}{\vec k}
\newcommand{\vp}{\vec p}
\newcommand{\vq}{\vec q}
\newcommand{\vkp}{\vec {k'}}
\newcommand{\vpp}{\vec {p'}}
\newcommand{\vqp}{\vec {q'}}
\newcommand{\bk}{{\bf k}}
\newcommand{\bp}{{\bf p}}
\newcommand{\bq}{{\bf q}}
\newcommand{\br}{{\bf r}}
\newcommand{\bR}{{\bf R}}
\newcommand{\up}{\uparrow}
\newcommand{\down}{\downarrow}
\newcommand{\fns}{\footnotesize}
\newcommand{\ns}{\normalsize}
\newcommand{\cdag}{c^{\dagger}}
\newcommand{\g}{\,\mbox{GeV}}

\title{A phenomenological study on the production of Higgs bosons in the cSMCS model at the LHC}

\author{$N.\;Darvishi^{a}$} \email{neda.darvishi@fuw.edu.pl} 
\author{$M.R. \; Masouminia^{b}$}\email{m.masouminia@ut.ac.ir}
\affiliation {$^{a}$ Faculty of Physics, University of Warsaw, Pasteura 5, 02-093 Warsaw, Poland
\\ $^{b}$ Department of Physics, University of $Tehran$, 1439955961, $Tehran$, Iran}

\begin{abstract}

In the present work, we intend to predict the production rates of the Higgs bosons in the simplest extension of the Standard Model (SM) by a neutral complex singlet (cSMCS). This model has an additional source of CP violation and provides strong enough first-order electroweak phase transition to generate the baryon asymmetry of universe (BAU). The scalar spectrum of the cSMCS includes three neutral Higgs particles with the lightest one considered to be the 125 GeV Higgs boson found at LHC. The SM-like Higgs boson comes mostly from the SM-like SU(2) doublet, with a small correction from the singlet. To predict the production rates of the Higgs bosons, we use a conventional effective LO QCD framework and the unintegrated parton distribution functions (UPDF) of Kimber-Martin-Ryskin (KMR). We first compute the SM Higgs production cross-section and compare the results to the existing theoretical calculations from different frameworks as well as the experimental data from the CMS and ATLAS collaborations. It is shown that our framework is capable of producing sound predictions for these high-energy QCD events in the SM. Afterwards we present our predictions for the Higgs boson production in the cSMCS. 

\end{abstract}

\maketitle

\section{Introduction}
\label{sec:intro}

Throughout the years numerous theoretical and phenomenological attempts have been made, trying to explore different aspects of the production of the Higgs particles at the LHC, within the Standard Model (SM), e.g. the references \cite{WattWZ, Lipatov, Luszczak:2005xs, Grazzini, Chen, Ferrera, Monni, Florian, Liebler, Caola,Telnov:2013zfa}. 
Here, we study the production of the Higgs bosons of the cSMCS (the extension of SM with a neutral complex singlet) ($h_1, h_2, h_3$) at the LHC, using $k_t$-factorization framework.
As it have been shown in \cite{Darvishi:2016gvm,Bonilla:2014xba, Darvishi:2016tni,Krawczyk:2015xhl,NDthesis}, the cSMCS model contains three neutral Higgs particles which the lightest is the $h_1=$125 GeV Higgs boson found at the LHC and the other two Higgs scalars, $h_i,\;i=2,3$ are taken to have masses 
$$
	M_{h_3} \gtrsim M_{h_2} > 150 \, {\rm GeV}.
$$

\begin{figure*}
\centering
\includegraphics[scale=0.4]{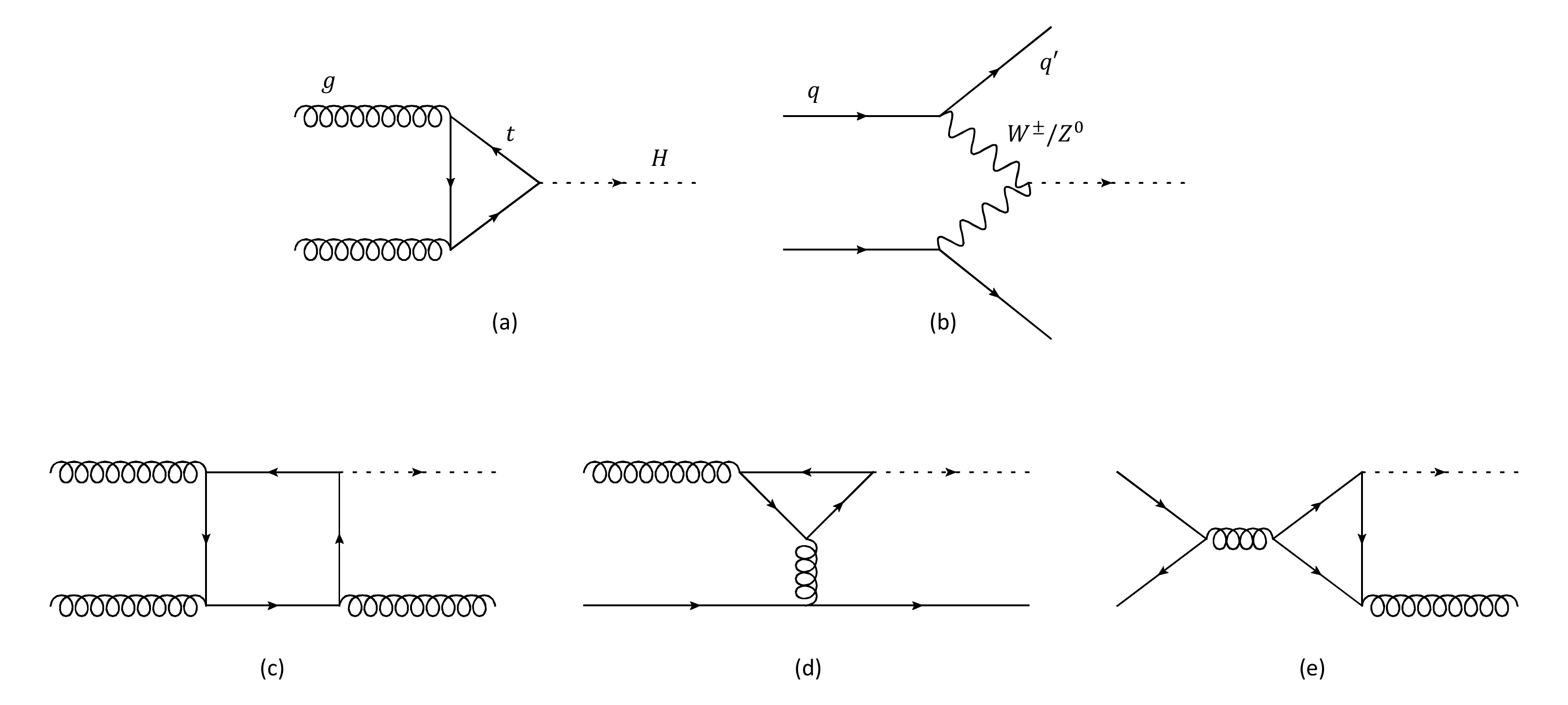}
\caption{The main contributing sub-processes in the total cross-section for the production of the Higgs bosons at the LHC.}
\label{fig1}
\end{figure*}

The main contribution to the cross-section for the Higgs bosons production at the LHC,
$$
	P_1 + P_2 \to H + X,
$$
gives the so-called gluon-gluon fusion sub-process, i.e. 
\begin{equation}
	g^{*}(k_1) + g^{*}(k_2) \to H(p),
	\label{eq1}
\end{equation}
see the figure \ref{fig1} part (a). Also, the Higgs boson production accompanied with a single jet or double jets can be traced back to the weak-boson fusion processes (figure \ref{fig1} part (b)) and $g^{*}+g^{*} \to H + g$, $g^{*}+q^{*} \to H + q$ and $q^{*}+\bar{q}^{*} \to H + g$ sub-processes (parts (c), (d) and (e) of the figure \ref{fig1}, respectively), which are expected to give roughly one tenth of the total Higgs production rate.
It has been shown that one can replace such complicated calculation by using a higher-order correction factor (i.e. the K-factor) \cite{WattWZ}. Also, it has been confirmed that using this K-factor will produce a good approximation of the full next-to-leading order (NLO) calculations (i.e. by counting the contributions of all the diagrams in the figure \ref{fig1}) \cite{NLO-Higgs}.   
 In principle, one has to include the contributions of all quark flavors in such diagrams. However, since the SM Higgs boson coupling to the top quark is considerably stronger compared to the other quarks, we consider only top-quark loops in our calculations. 


In the collinear factorization framework, the total cross-section for the production of a Higgs boson can be written as the partonic cross-section for the involving sub-process ($\hat{\sigma}_{gg \to H}$), times the probability of appearing that
particular partonic configuration at the top of the evolution ladder of the individual hadrons, i.e.
\begin{eqnarray}
  \sigma_{P+P \to H+X} &=&\int_0^1 {dx_1 \over x_1} \int_0^1 {dx_2 \over x_2}
  \; x_1 g(x_1,\mu_1^2)\; x_2 g(x_2,\mu_2^2) \; \nonumber \\[0.4cm]
  & \times & \hat{\sigma}_{gg \to H}(x_1,k^2_{1,t}=0,\mu_1^2;x_2,k^2_{2,t}=0,\mu_2^2). \nonumber \\
  \label{eq2}
\end{eqnarray}

The single-scaled (gluonic) parton distribution functions (PDF), $g(x_i,\mu_i^2)$, are the solutions of the Dokshitzer-Gribov-Lipatov-Altarelli-Parisi (DGLAP) evolution equations \cite{DGLAP1,DGLAP2,DGLAP3,DGLAP4}. These functions parametrize the probability of finding a gluon, emitting from the $i$th hadron and carrying the fraction $x_i$ of its longitudinal momentum. parameters $\mu_i$ are the ultra-violet cutoffs, related to the virtuality of the exchanged gluon during the inelastic scattering. In the Eq.(\ref{eq2}), $k_{i,t}$ are the transverse momenta of the incoming gluons. Neglecting the transverse momentum contributions of the incoming partons can seriously lower the precision of the calculations, predominantly for the event with the high center-of-mass energy in the small-$x$ region \cite{KMR, MRW, KKMS, KIMBER, WattWZ}. Knowing this have brought up the necessity of introducing transverse momentum dependent parton distribution functions (TMD PDF), notably the Ciafaloni-Catani-Fiorani-Marchesini (CCFM) evolution equation \cite{CCFM1,CCFM2,CCFM3,CCFM4,CCFM5}, and the Balitski-Fadin-Kuraev-Lipatov (BFKL) evolution equation \cite{BFKL1,BFKL2,BFKL3,BFKL4,BFKL5}. Other approach based on the unintegrated parton distribution functions (UPDF) with $k_t$-factorization are the leading order (LO) Kimber-Martin-Ryskin (KMR) and NLO Martin-Ryskin-Watt (MRW) formalisms \cite{KMR,MRW}. Recently, it has been shown that these UPDF, specially in the KMR formalism, provide successful descriptions of the existing high energy experimental data \cite{FL,Dipole-FL,NLO-W/Z,W/Z-13,Dijet}. 

{ {Several analysis have been previously published, addressing the production of the Higgs boson in the SM, using different visualizations of the $k_t$-dependent PDF, e.g. see \cite{WattWZ,Lipatov,Luszczak:2005xs}. In the present work, we calculate the total cross-section for production of the Higgs particles in the cSMCS within the framework of $k_t$-factorization, using the UPDF of KMR. By this end, first we calculate the total cross-section for the production of $H_{SM}$ particle and compare the results with the existing SM predictions from other theoretical analysis and also with the experimental data from the CMS and the ATLAS collaborations \cite{CMS8H, ATLAS8H, ATLAS13H}. In this part of the calculations, we follow the procedure of reference \cite{Lipatov}, with some technical differences in kinematic boundaries, energy scale and the choice of hard-scale (see section III). Afterward, we present our prediction for the production of the cSMCS Higgs particles (i.e. $h_1, h_2$ and $h_3$).
In our calculations, we have used the benchmarks presented in the reference  \cite{Darvishi:2016gvm}.}}

{ {The outlook of this paper is as follows: In section \ref{sec:smcs}, we will briefly introduce the cSMCS as a simple extension of the SM with a complex singlet scalar field. The potential of the model and its physical states will be discussed followed by presenting the benchmarks of cSMCS. In the section \ref{sec:Prod}, the $k_t$-factorization production of the Higgs boson will be reviewed. We will present some technical points regarding our numerical analysis in the section \ref{sec:Numeric} and discuss our results in the section \ref{sec:Results}. Our conclusion will be presented in the section \ref{sec:Conclusions}.}}

\section{The cSMCS: The SM plus a complex singlet scalar}
\label{sec:smcs}

The full Lagrangian of this model is given by
\begin{equation}
{ \cal L}={ \cal L}^{SM}_{ gf } +{ \cal L}_{scalar} + {\cal L}_Y(\psi_f,\Phi), 
\label{lagrbas}
\end{equation}
where the first term, ${\cal L}^{SM}_{gf}$, describes the interaction between SM gauge boson $(W^{\pm},Z)$-SM fermion, ${\cal L}_{scalar}$ describes the scalar sector of the model with one SU(2) doublet $\Phi$ and a neutral complex scalar (spinless) singlet $\chi$. ${\cal L}_Y(\psi_f,\Phi)$ represents the Yukawa interaction of $\Phi $ with SM fermions. Within our model, the neutral complex singlet $\chi$ does not couple to the SM fermions and gauge bosons. The singlet-SM fermion interactions are presented through the mixing of the singlet $\chi$ with the doublet $\Phi$ (it is the same for the singlet interaction with the gauge bosons). 
We assume $\Phi$ and $\chi$ fields to have non-zero vacuum expectation values ($vev$) $v$ and $w e^{i\xi}$, respectively ($v,w,\xi\in \bf{R}$). The following field decomposition around the vacuum state are used,
\begin{equation}
\Phi = \left( \begin{array}{c} \phi^+ \\ {1\over \sqrt{2}}(v + \phi_1 + i \phi_4) \\ \end{array} \right), 
\chi = {1\over \sqrt{2}}( w e^{i \xi} + \phi_2 + i \phi_3). \label{dec_singlet}
\end{equation}
Throughout this work, we use $w_1=w\cos\xi$ and $w_2=w\sin\xi$ definitions.
Masses of the gauge bosons are given by the $vev$ of the  doublet as in the SM, e.g $M_W^2 = g^2 v^2/4$ for the $W$ boson.
\subsection{Potential}
The scalar potential of the model can be written as follows  \cite{Darvishi:2016tni,Darvishi:2016gvm,Krawczyk:2015xhl,Bonilla:2014xba}
\begin{equation}
V=V_{D}+V_S+V_{DS}. \label{potgen}
\end{equation}
$V_{D}$ and $V_{S}$ are respectively the pure doublet and the pure singlet parts. The SM part of the potential, represented by $V_{D}$, is equal to
\begin{eqnarray}
V_{D} =&& -\frac{1}{2}{m_{11}^2} \Phi^\dagger\Phi 
+ \frac{1}{2}\lambda \left(\Phi^\dagger\Phi\right)^2.
\label{potSM}
\end{eqnarray}
The potential for a complex singlet $V_{S}$ is
\begin{eqnarray}
V_{S} =&& -\frac{1}{2}m_4^2 (\chi^{*2} + \chi^2) -\frac{1}{2} m_s^2 \chi^* \chi + \lambda_{s1}(\chi^*\chi)^2
\nonumber\\&&
 + \lambda_{s2} (\chi^*\chi)(\chi^2+\chi^{*2}) + \lambda_{s3} (\chi^4 + \chi^{*4})
\nonumber\\&&
+ \kappa_1 (\chi + \chi^*) + \kappa_2 (\chi^3 + \chi^{*3}) + \kappa_3(\chi+ \chi^*)(\chi^*\chi). \nonumber\\
\label{potS}
\end{eqnarray}
The doublet-singlet interaction terms are:
\begin{eqnarray}
V_{DS} =&&\Lambda_1(\Phi^\dagger\Phi)(\chi^* \chi) + \Lambda_2 (\Phi^\dagger\Phi)(\chi^2+\chi^{*2})
\nonumber\\&&
+ \kappa_4 (\Phi^\dagger\Phi) ( \chi +\chi^*). 
\end{eqnarray}
There are three quadratic terms ($m^2_i$), six dimensionless quartic ($\lambda_i, \Lambda_i$) and four dimensionful parameters $\kappa_{i},\;i=1,2,3,4$, describing linear ($\kappa_1$) and cubic terms ($\kappa_2,\kappa_3$ and $\kappa_4$). The linear term $\kappa_1$ can be removed by a translation of the singlet field. The potential is symmetric under a $\chi \to \chi^*$ transformation.
We impose a global $U(1)$ symmetry to reduce the number of parameters in the potential \cite{Darvishi:2016tni,Darvishi:2016gvm,Krawczyk:2015xhl,Bonilla:2014xba}.
\begin{equation}
U(1): \;\; \Phi \to \Phi,\, \chi \to e^{i\alpha} \chi .\label{u1def}
\end{equation}
{ {However, having a non-zero $vev$ for the singlet results in a spontaneous breaking of this U(1) symmetry and the appearance of mass-less Nambu-Goldstone scalar particles. This cannot be acceptable. To prevent this, we can consider a potential that has some U(1) soft-breaking terms. This means that the singlet cubic terms $\kappa_{2,3}$, $\kappa_4$ and the singlet quadratic term $m_4^2$ are kept. Therefore, the U(1)-symmetric terms, $m_{11}^2, m_{s}^2, \lambda, \lambda_{s1}$ and $ \Lambda_{1}$ and U(1)-soft-breaking terms $m_{4}^2$ and $ \kappa_{2,3,4}$ remain in the potential. For simplicity, we will use the following notation: $ \lambda_s= \lambda_{s1}, \Lambda=\Lambda_1$. The potential is as follows:}}
\begin{eqnarray} \label{potchi}
V =&& -\frac{1}{2}{m_{11}^2} \Phi^\dagger\Phi+ \frac{1}{2}\lambda\left(\Phi^\dagger\Phi\right)^2 + \Lambda(\Phi^\dagger\Phi)(\chi^* \chi)
\nonumber\\&&
 - \frac{1}{2}{m_4^2} (\chi^2+\chi^{*2})- \frac{1}{2}{m_s^2} \chi^* \chi+ \lambda_{s} (\chi^*\chi)^2
\nonumber\\&&
+ \kappa_2 (\chi^3 + \chi^{*3}) + \kappa_3 (\chi + \chi^*)(\chi^*\chi)
\nonumber\\&&
+ \kappa_4 (\Phi^\dagger\Phi) ( \chi +\chi^*). 
\label{potVS}
\end{eqnarray}
All the parameters of the potential are real. $V$ is also explicitly CP conserving. We shall call the model with this choice of parameters, cSMCS \cite{Darvishi:2016tni,Darvishi:2016gvm,Krawczyk:2015xhl}. 
\subsection{Physical states in the Higgs sector}
\label{physical state}
The mass matrix $M^2_{mix}$  that describes the singlet-doublet mixing, in the basis of neutral fields $\phi_1,\;\phi_2,\;\phi_3 $ is as follows:
\begin{equation}
M^2_{mix} = \left(
\begin{array}{ccc}
M_{11} & M_{12} & M_{13}\\
M_{21} & M_{22} & M_{23}\\
M_{31} & M_{32} & M_{33}
\end{array} \right),
\label{mixed}
\end{equation}
where the $M_{ij} (i,j=1,2,3)$ are:
\begin{eqnarray}
M_{11}=&& v^2 \lambda_1,
\nonumber\\ 
M_{12}=&&v (w_1\Lambda+{{\sqrt{2} \kappa_4}}), 
\nonumber\\ 
M_{13}=&&v w_2 \Lambda, 
\nonumber\\ 
M_{22}=&&\frac{w^2}{\sqrt {2} w_1} \left(3\kappa_2 +\kappa_3(1+2 (w_1^2-w_2^2)/w^2)\right.\nonumber\\ 
&&\left. -\kappa_4 v^2/w^2\right)+ 2 w_1^2 \lambda_s,
\nonumber\\ 
M_{23}=&&w_2(2 w_1 \lambda_s +\sqrt{2}(-3\kappa_2+\kappa_3)),
\nonumber\\ 
M_{33}=&& 2 w_2^2 \lambda_s.
\label{masses2}
 \end{eqnarray}

Diagonalization of $M^2_{mix}$, Eq.(\ref{mixed}), gives the mass-eigenstates { {of}} $h_1,h_2$ and $h_3$,
\begin{eqnarray}
	\left( \begin{array}{c} h_1\\ h_2\\ h_3\\ \end{array} \right) 
	&&  = r \left( \begin{array}{c} \phi_1\\ \phi_2\\ \phi_3\\ \end{array} \right), 
	\nonumber \\
	r M_{mix}^2 r^T &&  = diag(m_{h_1}^2,m_{h_2}^2,m_{h_3}^2),
	\label{diag}
\end{eqnarray}
where
\begin{eqnarray}
r = \left(
\begin{array}{ccc}
c_1 c_2 & c_3 s_1 - c_1 s_2 s_3 & c_1 c_3 s_2 + s_1 s_3\\
-c_2 s_1 & c_1 c_3 + s_1 s_2 s_3 & -c_3 s_1 s_2 + c_1 s_3\\
-s_2 & -c_2 s_3 & c_2 c_3
\end{array} \right).\label{rotfull}
\end{eqnarray}
All $\alpha_i$ vary over an interval of length $\pi$. The full rotation matrix $r$ depends on the mixing of $\alpha_1, \alpha_2$ and $ \alpha_3$ angles and  $c_i = \cos \alpha_i $ and $ s_i = \sin \alpha_i$.
An important relations can be derived from the equations (\ref{diag}) and (\ref{rotfull}), i.e,
\begin{equation}
\phi_1 = c_1 c_2 h_1 - c_2 s_1 h_2 - s_2 h_3, \label{phi1DM}
\end{equation}
with $r_{11}=c_1 c_2$,  $r_{21}=c_2 s_1$ and $r_{31}=s_2$.
The Yukawa interactions between the Higgs fields and the fermions ($\mathcal{L}_Y$) generates the fermion masses. Note that only the doublet couples to the fermions,
\begin{equation}
  \mathcal{L}_Y = -\sum_f \frac{m_f}{v}\overline ff(r_{11}h_1+r_{21}h_2+r_{31}h_3).
\end{equation}

{ {The couplings of the lightest Higgs particle ($h_1$) to the quarks and the gauge bosons in the cSMCS model, as compared with the corresponding couplings of the SM Higgs, are modified (suppressed) by a factor $r_{11}$.
Decay of the Higgs bosons into vector bosons ($V=Z,W$) is given by:
 \begin{equation}
 \Gamma(h_{1} \to VV^{*})= r_{11}^{2} \Gamma(H_{SM} \to VV^{*} ).
 \end{equation}
On the other hand, the main contribution in the one-loop coupling of $h_1$ to photons is due to
 the W boson and top quark, and therefore in the cSMCS, the corresponding amplitude and the decay rate are equal to:
\begin{eqnarray}
A(h_1\to \gamma \gamma) &=& r_{11} ( A^{SM}_W+ A^{SM}_t)  \nonumber \\
\to \Gamma (h_1\to \gamma \gamma) &=& r_{11}^2 \Gamma (H_{SM} \to \gamma \gamma).
\end{eqnarray}
The decay width of the Higgs $h_1$ into gluons is given by:
 \begin{equation}
 \Gamma(h_1 \to gg)= r_{11}^2 \Gamma(H_{SM} \to gg ).
 \end{equation}
Also, since the total width of the light Higgs boson $h_1$ is given by
 \begin{equation}
 \Gamma_{tot}\approx r_{11}^{2} \Gamma_{tot}^{SM}.
 \end{equation}
The signal strength $R_{\gamma\gamma}$ is given by,
\begin{eqnarray}
\label{Rxx}
 \mathcal{R}_{\gamma \gamma} =&& \frac{ \sigma( gg\to h_1 ) }{ \sigma( gg\to H_{SM} ) } 
  \frac{\text{BR}(h_1 \to \gamma \gamma) }{\text{BR} (H_{SM} \to\gamma \gamma) }
\nonumber\\  
=&&
 \frac{ \Gamma(h_1 \to gg) }{ \Gamma(H_{SM} \to gg ) } \, 
  \frac{\text{BR}(h_1 \to \gamma \gamma)}{\text{BR}(H_{SM} \to \gamma \gamma)},
\end{eqnarray}
taking into account that the production of the Higgs bosons in the LHC is dominated by the gluon fusion processes and that
the narrow width approximation can be applied.
The signal strengths $\mathcal{R}_{VV}$ and $\mathcal{R}_{\gamma \gamma}$ have the same amounts and are given by,
\begin{equation}\label{rgc}
\mathcal{R}_{VV}=\mathcal{R}_{\gamma \gamma}\approx r_{11}^2.
\end{equation}
In all benchmarks, ${r}_{11}^2 \sim 0.81-0.98$, which is in agreement with the LHC data on the 125 GeV Higgs couplings to $ZZ$.
The benchmarks of the cSMCS from reference \cite{Darvishi:2016gvm} are presented in the table \ref{tab-1}, including their corresponding $\mathcal{R}_{\gamma \gamma}$, $S$ and $T$ parameters.{{
We have also checked the validity of above benchmarks, with respect to the existing experimental bounds, using the $\mathtt{HiggsBounds}$ package (version 4.3.1) \cite{Bechtle:2008jh}, which includes constraints from direct Higgs searches at the LEP, the Tevatron and the LHC and selects the most sensitive exclusion limit for each parameter (at 95\% C.L.).}}

\begin{table*}
\footnotesize
\centering
\begin{tabular}{|c||c|c|c|c|c|c|c|c|c|c|c|c|c|}
\hline
 BP & $\alpha_1$ &$\alpha_2$&$\alpha_3$&$M_{h_{1}}$ &$M_{h_{2}}$&$M_{h_{3}}$&$S$&$T$&$R_{\gamma\gamma}^{h_{1}}$\\ \hline
 \hline 
 $A1$ &-0.047 &-0.053 &1.294 & 124.64 & 652.375 & 759.984 & -0.072 & -0.094 &0.98 \\ \hline 
 $A2$ &-0.048 &0.084 &0.084 & 124.26 & 512.511 & 712.407 & -0.001 & -0.039 & 0.98 \\ \hline 
 $A3$ &0.078 &0.297 &0.364 & 124.27 & 582.895 & 650.531 & 0.003 & -0.046 &0.98 \\ \hline 
 $A4$ &0.006 &-0.276 &0.188 & 125.86 & 466.439 & 568.059 & -0.013 & -0.169 &0.92 \\ \hline
 $A5$ &0.062 &-0.436 &0.808 & 125.21 & 303.545 & 582.496 & 0.002 & -0.409 &0.81 \\ \hline
\end{tabular}
\caption{Benchmark points $A1-A5$. Masses are given in GeV. The signal strength of the SM-like Higgs boson into $ \gamma\gamma$ is from ATLAS \cite{ref1} ($1.14^{+0.27}_{-0.25}$) and CMS \cite{ref2} ($1.11^{+0.25}_{-0.23}$). The latest values of the oblique parameters, determined from a fit with reference mass-values of top and Higgs boson $M_{t,ref}=173 \g$ and $M_{h,ref}=125 \g$ are $S = 0.05\pm0.11, T = 0.09\pm0.13$ \cite{ref3}.} 
\label{tab-1}
\end{table*}
 }}

In the next section, we will introduce the necessary theoretical framework for the computation of the SM and cSMCS Higgs production rates in the high-energy QCD hadron-hadron collisions.

\section{Calculation of the Higgs production cross-section }
\label{sec:Prod}

{ {Assuming that the gluons entering the $g^{*}+g^{*} \to H$ sub-process have some non-negligible transverse momenta, the total cross-section for the Higgs particle production can be calculated, using the definition of the $k_t$-factorization \cite{kt-fact}, which is given by
\begin{equation}
  a(x,\mu^2) = \int^{\mu^2} {dk_t^2 \over k_t^2} f_a(x,k_t^2,\mu^2),\label{eq3}
\end{equation}
where $a(x,\mu^2)$ represents the solutions of the DGLAP evolution equations for both quarks and qluons, i.e. $xq(x,\mu^2)$ and $xg(x,\mu^2)$ respectively.}} Here, the $f_a(x,k_t^2,\mu^2)$ are the corresponding UPDF of KMR formalism (for more description of the structure and the kinematics of the UPDF see the references \cite{KMR,MRW,NLO-W/Z}). Thus the Eq.(\ref{eq2}) can be rewritten as follows:
\begin{eqnarray}
	\sigma_{P+P \to H+X} && = \int_0^1 {dx_1 \over x_1} \int_0^1 {dx_2 \over x_2}
	\int_{0}^{\infty} {dk^2_{1,t} \over k^2_{1,t}} \int_{0}^{\infty} {dk^2_{2,t} \over k^2_{2,t}}
	\nonumber \\ && \times 
	f_g(x_1,k^2_{1,t},\mu_1^2) f_g(x_2,k^2_{2,t},\mu_2^2) 
	\nonumber \\ && \times
	\hat{\sigma}_{gg \to H}(x_1,k^2_{1,t},\mu_1^2;x_2,k^2_{2,t},\mu_2^2).
  \label{eq4}
\end{eqnarray}
The partonic cross-section, $\hat{\sigma}_{gg \to H}$, is defined as
\begin{equation}	
	d\hat{\sigma}_{gg \to H} = {d\phi_{gg \to H} \over F_{gg \to H}}
	|{\mathcal{M}}(g^{*}(k_1)+g^{*}(k_2) \to H(p))|^2,
  \label{eq5}
\end{equation}
where $k_i$ and $p$ respectively represent the 4-momenta of the incoming gluons and the produced Higgs boson. $d\phi_{gg \to H}$ and $F_{gg \to H}$ are the corresponding particle phase space and the flux factor,
	\begin{equation}	
	d\phi_{gg \to H} = {d^3 p \over 2E} \delta^{(4)} \left( k_1 + k_2 - p \right),
  \label{eq6}
	\end{equation}
	\begin{equation}	
	F_{gg \to H} = x_1 x_2 s,
  \label{eq7}
	\end{equation}
with $s$ being the center of mass energy squared,
	$$ s=(p_1 + p_2)^2=2p_1 \cdot p_2, $$
in the infinite momentum frame (where one can safely neglect the masses of the incoming hadrons in comparison with their momenta ($p_i \gg m_{i}$)). The $d\phi_{gg \to H}$ can be expressed in terms of the transverse momenta of the produced Higgs boson $p_{t}$, its rapidity $y_H$, and the azimuthal angles of its emission, $\varphi$,
	\begin{equation}	
	{d^3 p \over 2E} = {\pi \over 2} \; dp_{t}^2 \; dy_H \; {d\varphi \over 2\pi}.
  \label{eq8}
	\end{equation}
In the Eq.(\ref{eq5}), ${\mathcal{M}}$ is the matrix element of the $g^{*}(k_1)+g^{*}(k_2) \to H(p)$ sub-process, equal to (see the Appendix \ref{AppB}):
\begin{eqnarray}
	|\mathcal{M}|^2 = {\alpha_S^2(\mu^2) \over 288\pi^2} {G_F \over \sqrt{2}}
	 \; \tau^2 \; |D(\tau)|^2 \; ( m_H^2 + {p}_t^2 )^2 \; cos^2 \varphi.
	 \nonumber \\
	\label{eq11}
\end{eqnarray}

In a high-energy inelastic collision at the LHC, one can consider the following kinematics in the center-of-mass frame
$$
	p_i = {\sqrt{s} \over 2} (1,0,0,\pm 1),
$$
\begin{equation}
	\textbf{k}_i = x_i \textbf{P}_i + \textbf{k}_{i,\perp}, \;\;\; \textbf{k}_{i,\perp}^2 = -k_{i,t}^2, \;\;\; i=1,2 \;,
  \label{eq14}
\end{equation}
and express the law of the transverse momentum conservation for the $g^{*}(k_1) + g^{*}(k_2) \to H(p)$ sub-process as:
\begin{equation}
	\textbf{k}_{1,\perp} + \textbf{k}_{2,\perp} = \textbf{p}_{\perp},
  \label{eq15}
\end{equation}
with $\textbf{p}^2_{\perp} = - p_t^2$ being the transverse momentum of the produced Higgs boson. The longitudinal fractions $x_i$ can be expressed by the transverse mass of the Higgs boson, $m_{H,t}^2 \equiv m_H^2 + p_t^2$, its rapidity and the parameter $s$,
\begin{eqnarray}
	x_1 && = {m_{H,t} \over \sqrt{s}} e^{+y_H},
	\nonumber \\
	x_2 && = {m_{H,t} \over \sqrt{s}} e^{-y_H}.
  \label{eq16}	
\end{eqnarray}

Putting the above formulas together, we derive the master equation for the production of the Higgs bosons:
\begin{eqnarray}
	\sigma_{P+P \to H+X} && = {G_F \over \sqrt{2}} \int  
	{dk_{1,t}^2 \over k_{1,t}^2} \; {dk_{2,t}^2 \over k_{2,t}^2} \; dy_H \; {d\varphi \over 2\pi} \;
	cos^2 \varphi 
	\nonumber \\ && \times
	f_g(x_1,k_{1,t}^2,\mu^2) \; f_g(x_2,k_{2,t}^2,\mu^2) 
	\nonumber \\ && \times
	{\alpha_S^2(\mu^2) \over 144\pi} {\tau^2 \; |D(\tau)|^2 \over x_1 x_2 s m_H^2 } 
	\left( m_H^2 + {p}_t^2 \right)^2.
  \label{eq17}
\end{eqnarray}
To determine the density functions of the incoming gluons, $f_g(x_i,k^2_{i,t},\mu^2)$, we utilize the KMR formalism and obtain
\begin{eqnarray}
  f_g(x,k_t^2,\mu^2) && = T_g(k_t^2,\mu^2) {\alpha_S(k_t^2) \over 2\pi} \int^{z_{max}}_{x} dz 
  \nonumber \\ && \times
  [ 
  P_{gq}^{(LO)}(z) \sum_q {x \over z} q\left( {x \over z}, k_t^2 \right) 
	\nonumber \\ &&  
  + P_{gg}^{(LO)}(z) {x \over z} g\left( {x \over z}, k_t^2 \right) 
  ].
  \label{eq18}
\end{eqnarray}
The variable $z_{max}=\mu /(\mu + k_t)$ describes the angular ordering constraint, as a consequence of the color coherence effect of successive gluonic emissions \cite{AOC}. $T_g(k_t^2,\mu^2)$ is the probability of survival, which limits the parton emissions between the scales $k_t$ and $\mu$. It factors over the virtual contributions from the gluonic LO DGLAP equation and can be defined as:
\begin{eqnarray}
  T_g(k_t^2,\mu^2) && = exp [ - \int_{k_t^2}^{\mu^2} {\alpha_S(k^2) \over 2\pi}
  {dk^{2} \over k^2} \int^{z_{max}}_{0} dz
	\nonumber \\ && \times 
  \left( P_{gg}^{(LO)}(z) + n_f P_{gq}^{(LO)}(z) \right) ], 
  \label{eq19}
\end{eqnarray}
with $n_f$ being the number of active quark flavors. $P_{ab}^{(LO)}(z)$ are the LO splitting functions, parameterizing the probability of a parton with the longitudinal momentum fraction $x$ to be emitted from a parent parton with the fraction $x'$, $z=x/x'$ \cite{Dijet,PNLO}.

In the following section, we will introduce some of the numerical methods that have been used in the calculation of the Eq.(\ref{eq17}) to predict the total cross-section for the production of the Higgs bosons of the cSMCS, with the lightest being the SM-like Higgs boson. 

\section{Numerical analysis}
\label{sec:Numeric}

We use the UPDF of KMR, Eq.(\ref{eq18}), to numerically solve the Eq.(\ref{eq17}), utilizing the $\mathtt{VEGAS}$ algorithm in Monte-Carlo integration \cite{VEGAS}. The required PDF for the preparation of these UPDF are provided in the form of libraries, e.g. the MMHT2014 libraries, the reference \cite{MMHT2014}, where the single-scaled solutions of the DGLAP evolution equations have been fitted to the experimental data on the $F_2$ structure function from e-p deep inelastic scattering and the high-energy hadron-hadron scatterings. We chose the hard-scale of the UPDF as the transverse mass of the produced Higgs boson, i.e. 
\begin{equation}
	\mu = (m_{H}^2 + p_{t}^2 )^{1/2}.
	\label{eq20}
\end{equation}
One should note that the upper and the lower boundaries of the transverse momentum integrations in the Eq.(\ref{eq17}) are respectively $\infty$ and zero. Nevertheless, since the KMR UPDF rapidly converge to zero in the $k_t>\mu$ domain, it is safe to introduce an upper bound for these integrations in the following form
\begin{equation}
	k_{t,max} = \mu_{max} \equiv 4(m_{H}^2 + p_{t,max}^2 )^{1/2}.
	\label{eq21}
\end{equation}
Further domain has no influence on our results. On the other hand, it is important to note that the UPDF of the $k_t$-factorization are being defined only in the QCD perturbative regime, i.e. for $k_t>\mu_0$ with $\mu_0=1$ GeV, as the minimum scale for which the DGLAP evolution of the integrated PDF is valid. We have to define our treatment of these distribution functions in the non-perturbative region, $k_t<\mu_0$. A natural choice to by-pass this obstacle is to fulfill the requirement that 
\begin{equation}
	\lim_{k_{i,t}^2 \rightarrow 0} f_{g}(x_i,k_{i,t}^2,\mu^2) \sim k_{i,t}^2. \nonumber
\end{equation}
So, for the non-perturbative region, we choose
\begin{equation}
	f_g(x_i,k_{i,t}^2<\mu_0^2,\mu^2) = {k_{i,t}^2 \over \mu_0^2} x_i g(x_i,\mu_0^2) T_g(\mu_0^2,\mu^2).
  \label{eq22}
\end{equation}
Also, we set the boundaries of the rapidity integration in accordance with the specifications of the detectors (i.e. $|y_H| < 2.5$ for the CMS report \cite{CMS8H} and $|y_H| < 2.4$ for ATLAS reports \cite{ATLAS8H, ATLAS13H}, excluding the $1.37 < |y_H| < 1.52$ region for the later). Otherwise, we choose to integrate over the $|y_H| < 10$ domain. According to the Eq.(\ref{eq16}), and the $0<x_i<1$ constraint, further rapidity domain will have no influence on our result.

{ {At this point, we must mention that the higher order QCD corrections into the the total cross-section for production of the SM Higgs boson are non-negligible (see e.g. \cite{WattWZ,Lipatov,Luszczak:2005xs}). 
}} These correction are either kinematic in nature (e.g. corrections from the (b), (c), (d) and (e) diagrams in the figure \ref{fig1}) or arise from real parton emissions or virtual loop corrections. It is however customary to compensate for these neglected contributions by the means of introducing an additional factor into the main calculations, called the K-factor, which is defined as the ratio of the corrected results to the LO results. It has been suggested that introducing a K-factor as
\begin{equation}
	K = exp \left( C_A {\pi \alpha_S(\mu_c^2) \over 2} \right),
	\label{eq23}
\end{equation}
with $C_A=3$ and $\mu_c = ({ m_H p_t^2})^{1/3}$ can absorb the main part of these higher order corrections \cite{WattWZ}.

At this point, we are ready to calculate the cross-section for the production of the SM Higgs boson and the cSMCS Higgs bosons, using the master Eq.(\ref{eq17}) in the $k_t$-factorization framework. To switch from the SM to the cSMCS, we change the mass of the considered Higgs boson and replace the couplings of SM Higgs boson with quarks, $ m_f/v$, with its corresponding cSMCS couplings, i.e. $r_{i1} m_f/v$. We use the benchmarks presented in the table \ref{tab-1} for the mass of the considered Higgs boson. 

{{
As a final remark, it should be reemphasized that in these calculations we have followed the footsteps of the reference \cite{Lipatov} with the following technical differences:
\begin{itemize}
	\item[$\circ$] To capture the contributions of the incoming partons, we use the UPDF of KMR instead of the TMD PDF of the CCFM evolution equation.
	\item[$\circ$] We bound the rapidity regions, according to the specifications of the CMS and ATLAS reports. We also exclude the non-perturbative $k_t<1$ GeV domain. To compensate for the later exclusion, we utilize the prescription that is given as the Eq.(\ref{eq22}). 
	\item[$\circ$] Our calculations have been performed in $E_{CM}=8, \; 13$ and $14$ TeV while in \cite{Lipatov}, $E_{CM}$ is fixed to be $14$ TeV.
	\item[$\circ$] We have chosen the hard-scale of the process to be the transverse mass of the produced Higgs boson, see the Eq.(\ref{eq20}), in line with the procedure of the references \cite{NLO-W/Z,W/Z-13}. This is a more convenient choice, compared to $\mu=m_H$ that have been used in \cite{Lipatov}.
\end{itemize}
}}

\section{Results and discussions}
\label{sec:Results}
\begin{figure*}
\centering
\includegraphics[scale=0.3]{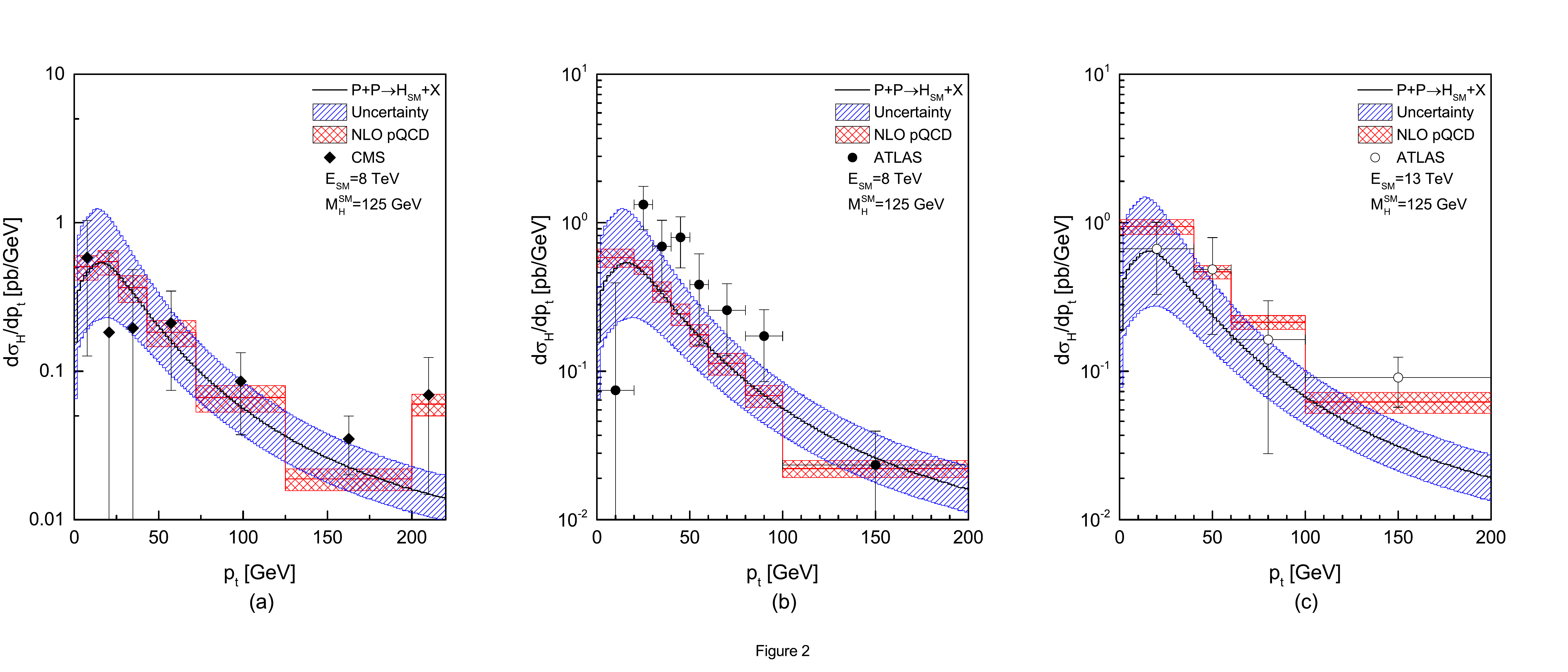}
\caption{Differential cross-section for the production of the SM Higgs boson as a function of the transverse momentum of the SM Higgs boson at the LHC. The calculations have been performed for the center-of-mass energies $E_{cm} = 8$\;TeV and $13$\;TeV. The black solid curves illustrate the main prediction while the blue stripped patterns show the uncertainty bound for the results. The uncertainty bounds are determined via manipulating the hard-scale $\mu$ by a factor of 2. The results have been compared with the experimental data from the CMS (a) and the ATLAS (b and c) collaborations \cite{CMS8H,ATLAS8H,ATLAS13H}. To prepare the KMR UPDF, we have utilized the PDF of MMHT2014. The collinear results (NLO pQCD) are from the \cite{CMS8H,ATLAS8H,ATLAS13H}.}
\label{fig2}
\end{figure*}

\begin{figure*}
\centering
\includegraphics[scale=0.3]{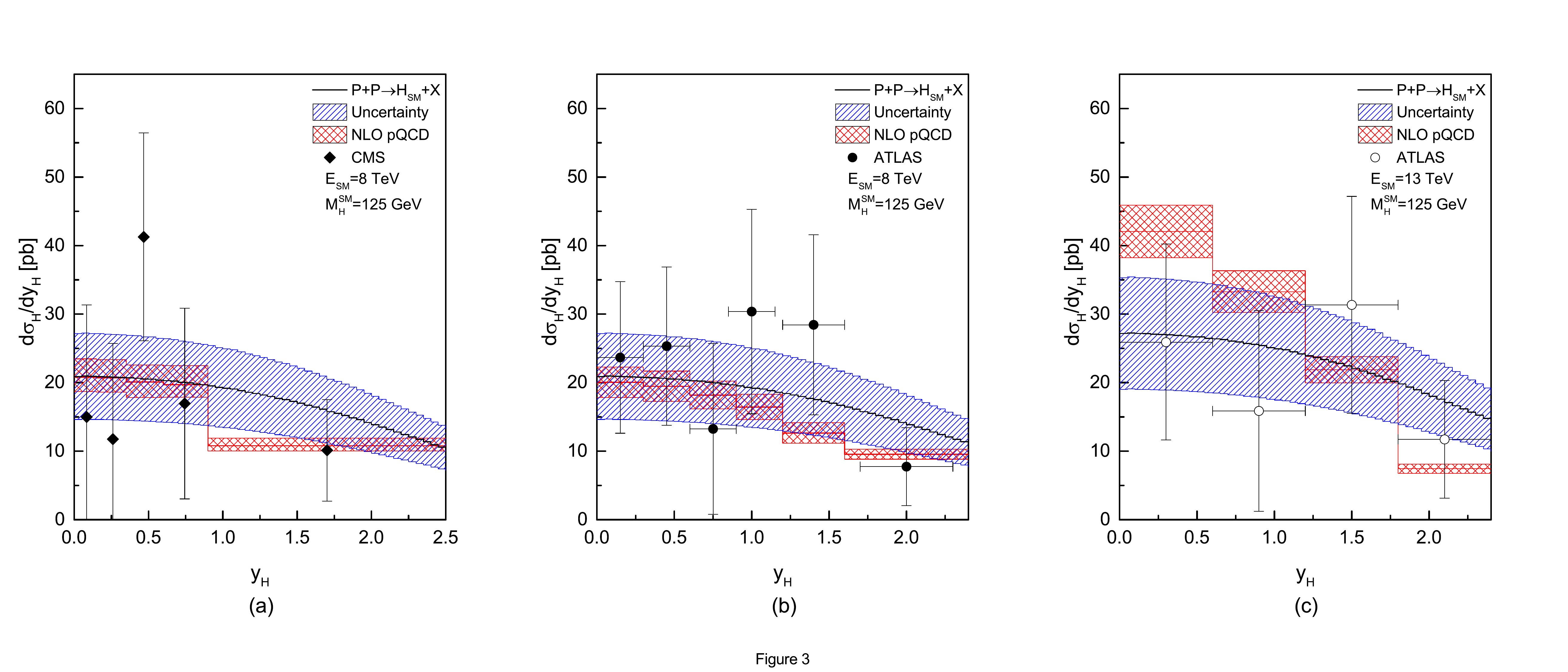}
\caption{Differential cross-section of the production of the SM Higgs boson at the LHC as a function of the rapidity of the produced particle. other details are as in the figure \ref{fig2}.}
\label{fig3}
\end{figure*}

Before discussing the cSMCS predictions, it is necessary to prove that our framework can in fact produce reliable results. We test our approach by calculating the cross-section for the production of the SM Higgs boson, $H_{SM}$. The figure \ref{fig2} presents the differential cross-section for the production of the SM Higgs boson ($d\sigma_H/dp_t$) versus the transverse momentum of the produced particle ($p_t$). Parts (a) and (b) illustrate our results for the LHC with the center-of-mass energy $E_{CM}=8$\;TeV and the rapidity regions $|y_H| < 2.4$ and $|y_H| < 2.5$, respectively. The main results are being presented by solid black curves while the blue stripped patterns mark the uncertainty bounds (which are determined by manipulating the hard-scale $\mu$ by a factor of 2). Part (c) shows the results for the center-of-mass energy $E_{CM}=13$\;TeV and rapidity region $|y_H| < 2.4$ (excluding the $1.37 < |y_H| < 1.52$ region). { {These panels also contain the collinear results of the NLO perturbative QCD (pQCD) calculations from \cite{CMS8H,ATLAS8H,ATLAS13H}.}}
The results are compared with CMS data at 8 TeV in the figure \ref{fig2}(a) and with ATLAS data at 8 TeV and 13 TeV in the figure \ref{fig2}(b) and \ref{fig2}(c), respectively. 
The data points are the results of measurements of the CMS and the ATLAS collaborations, references \cite{CMS8H,ATLAS8H,ATLAS13H}. 
A similar comparison is presented in the figure \ref{fig3}, regarding the rapidity contribution of the SM Higgs boson production. i.e. $d\sigma_H/dy_H$ versus $y_H$.

We conclude that our framework, at least within its uncertainty bounds, can give an acceptable description of the SM Higgs particle.

\begin{figure*}
\centering
\includegraphics[scale=0.3]{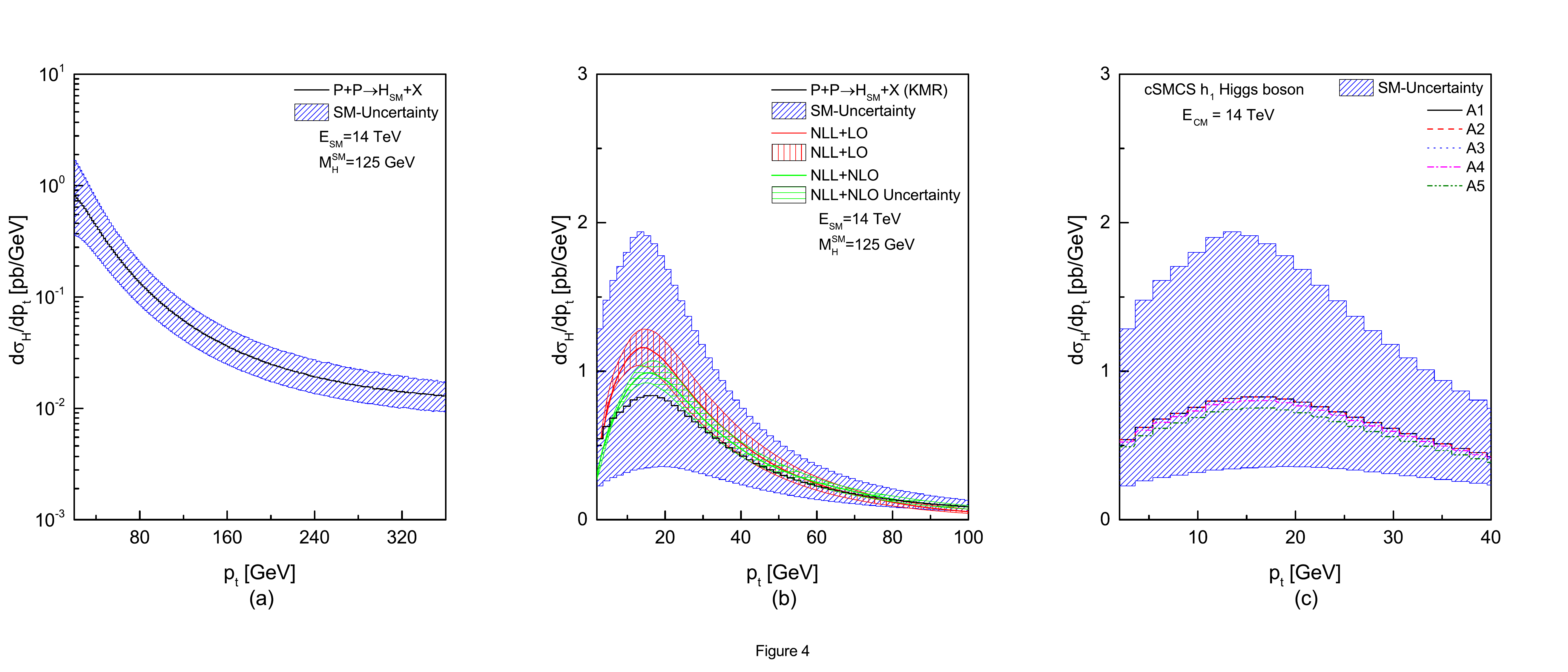}
\caption{Differential cross-section of the production of the Higgs boson at the LHC as a function of the transverse momentum at $E_{CM}=14$\;TeV. Part (a) illustrates the main results and the corresponding uncertainty bounds for the SM Higgs boson, $H_{SM}$. Part (b) presents a comparison between our results for $H_{SM}$ (the solid black curve and its blue strip-patterned uncertainty bounds) and the results of similar calculations within the collinear framework, i.e. the NLL+LO (red curve within the red uncertainty bounds) and the NLL+NLO (green curve within the green uncertainty bounds). The collinear results are from the reference \cite{Grazzini}. Part (c) presents the predictions of the various benchmarks of cSMCS for the lightest SM-like Higgs boson $h_{1}$.}
\label{fig4}
\end{figure*}

Figure \ref{fig4}, illustrates our predictions for differential cross-section for the production of the SM Higgs boson at the LHC for $E_{CM}=14$\;TeV, as functions of its $p_t$. Part (a) depicts the general behavior of this production rate, exclusively within the SM. Part (b) outlines an comparison between our calculations and the similar results from NLL+LO (next-to-leading logarithmic re-summation plus LO calculations) and NLL+NLO (next-to-leading logarithmic re-summation plus NLO calculations) analysis performed in the collinear factorization, in \cite{Grazzini}. The collinear results show a slightly higher peak, compared to the KMR framework. Otherwise, the general behavior of these frameworks are identical. In the part (c), we demonstrate the differential cross-section of the production of the SM-like Higgs boson $h_1$ from the cSMCS. The curves A1 through A5 correspond to the cSMCS benchmarks, the table \ref{tab-1}. These computations are presented with respect to the SM uncertainty bounds. 

\begin{figure*}
\centering
\includegraphics[scale=0.3]{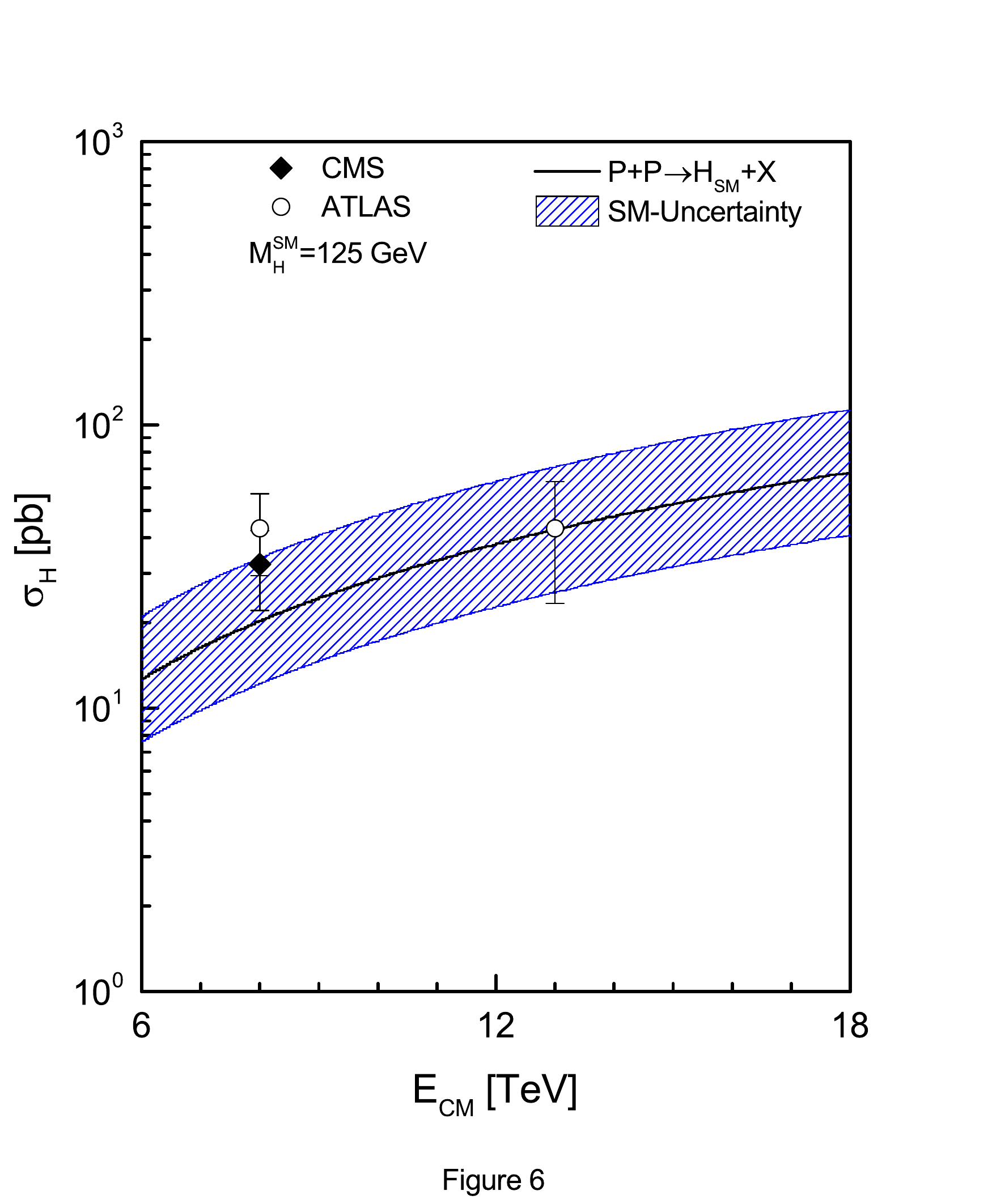}
\caption{Inclusive total cross-section of the production of the SM Higgs boson $H_{SM}$ at the LHC as a function of the central-mass energy of proton-proton collision, at the LHC. The solid black curve illustrates the main results while the uncertainty bounds (blue stripped pattern) have been produced via manipulating the hard-scale of the UPDF, $\mu$, by a factor of 2. The experimental data are from the CMS (black diamonds) and the ATLAS (white circles) collaborations \cite{CMS8H, ATLAS8H, ATLAS13H}.}
\label{fig6}
\end{figure*}

\begin{figure*}
\centering
\includegraphics[scale=0.3]{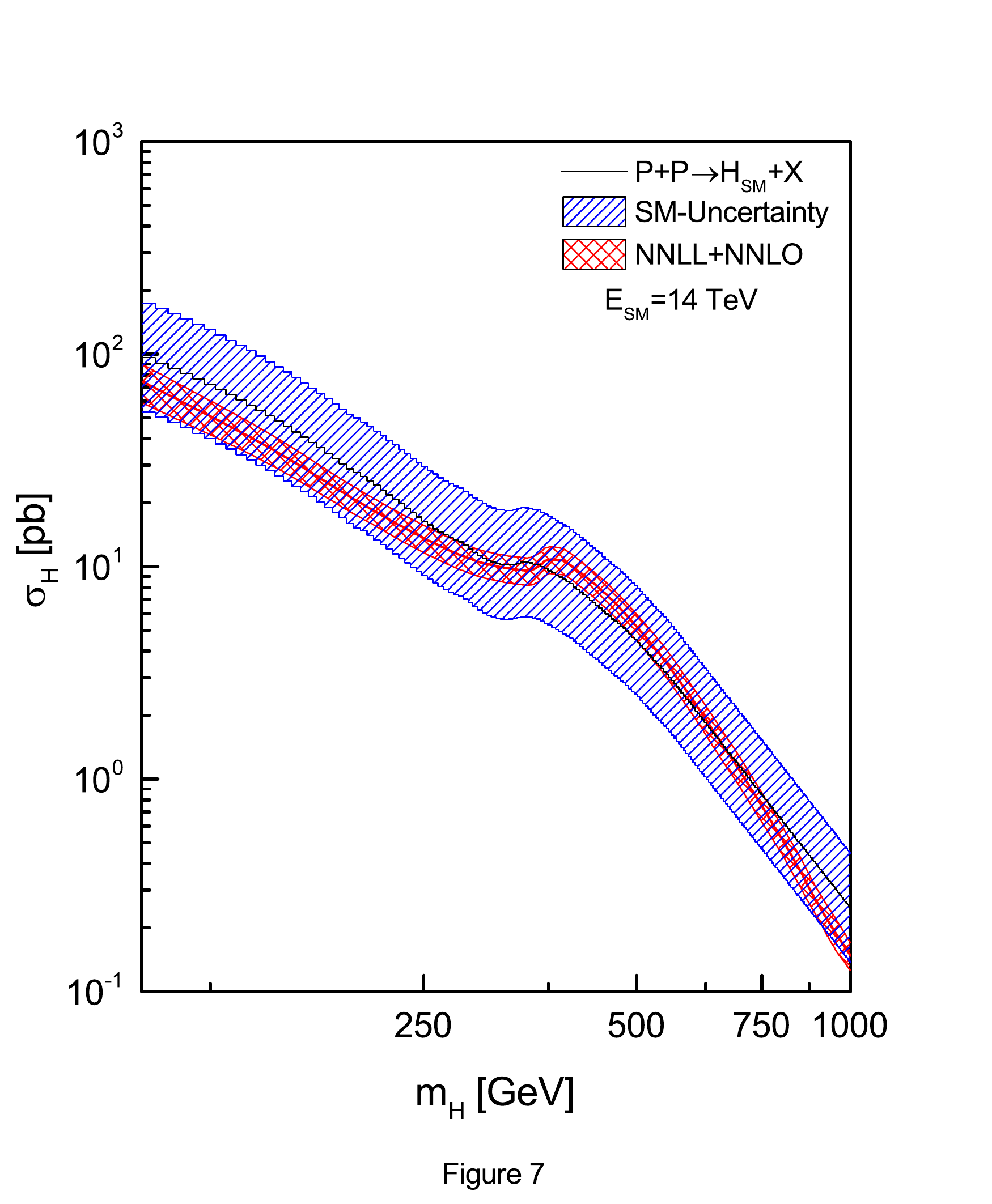}
\caption{Inclusive total cross-section of the production of the Higgs bosons $H_{SM}$ as a function of the mass at $E_{CM}=14$\;TeV. The solid black curve illustrates the main results while the uncertainty bounds (blue stripped pattern) have been produced via manipulating the hard-scale of the UPDF, $\mu$, by a factor of 2.  The collinear results (NNLL+NNLO) are from the reference \cite{Telnov:2013zfa}.}
\label{fig7}
\end{figure*}

Furthermore, we have calculated the total rate of production of the SM Higgs boson, $H_{SM}$, at the LHC ($\sigma_H$) as a function of the center-of-mass energy of the hadronic collision ($E_{CM}$) and compared the results with the existing experimental data from CMS and ATLAS \cite{CMS8H,ATLAS8H,ATLAS13H}, see the figure \ref{fig6}. Our predictions seem to be realistic. Furthermore, we have computed for $E_{CM}=14$\;TeV the mass distribution of the SM Higgs boson, what is presented as the figure \ref{fig7}. The peculiar pattern which is seen in the cross-section is originated from the imaginary part of $D(\tau)$, the Eq.(\ref{eq12}), which becomes non-zero at $m_H = 2m_t$. { {Figure \ref{fig7} also contains the collinear NNLL+NNLO (next-to-next-to-leading logarithmic re-summation plus next-to-leading calculations) results from the reference \cite{Telnov:2013zfa}. The comparison between two frameworks show relative agreement. }}

\begin{figure*}
\centering
\includegraphics[scale=0.3]{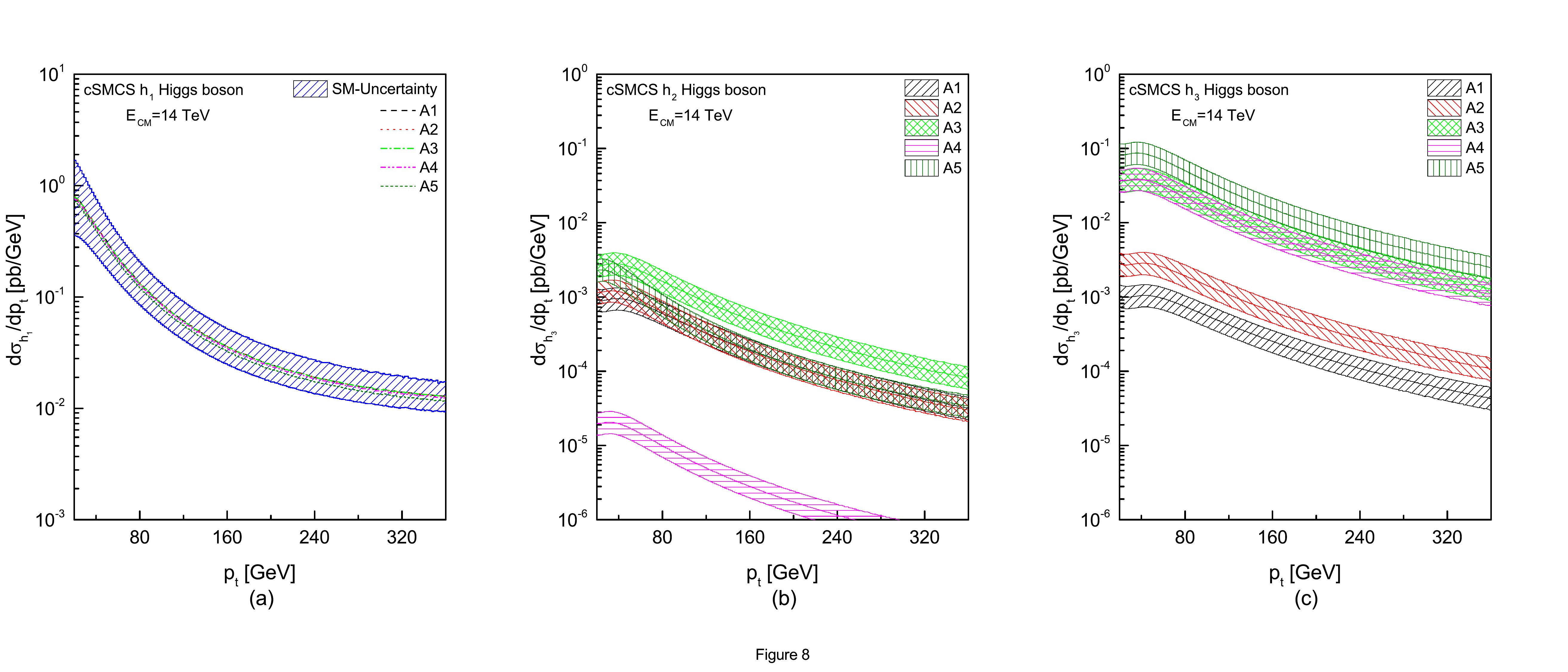}
\caption{Differential cross-section for the production of the cSMCS Higgs bosons at the LHC as a function of the transverse momentum of the produced Higgs bosons at $E_{CM}=$14\;TeV. Parts (a), (b) and (c) present the benchmark predictions of the cSMCS Higgs bosons, $h_1$, $h_2$ and $h_3$ in accordance with the benchmarks given in the table \ref{tab-1}. To calculate the uncertainty bound, the hard-scale $\mu$ has been manipulated by a factor of 2.}
\label{fig8}
\end{figure*}

\begin{figure*}
\centering
\includegraphics[scale=0.3]{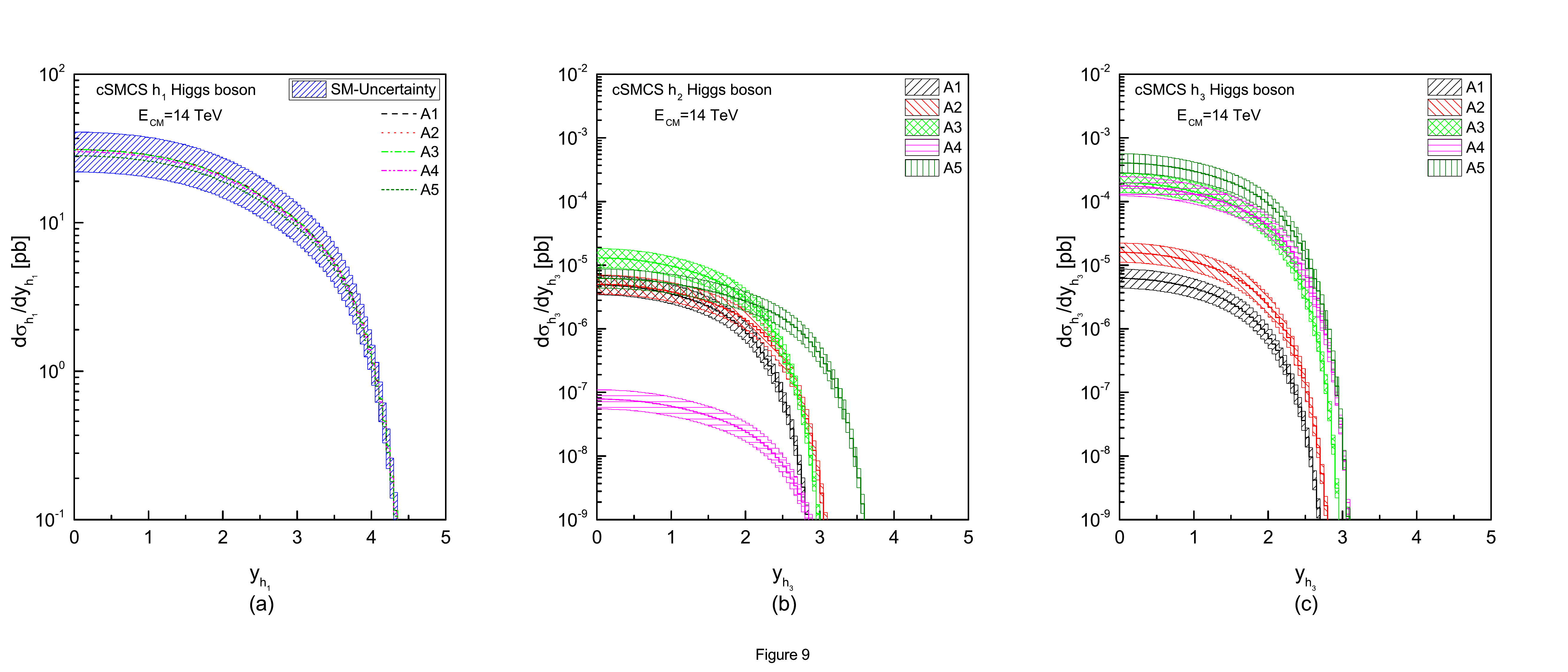}
\caption{Differential cross-section for the production of the cSMCS Higgs bosons at the LHC as a function of the rapidity of the produced Higgs particles at $E_{CM}=14$\;TeV. Other details are as in the figure \ref{fig8}.}
\label{fig9}
\end{figure*}

At this point, after proving our approach in describing the SM Higgs production at the LHC, we present our predictions regarding the expected production rate for { {$h_1$}} and for the heavier Higgs bosons $h_2$ and $h_3$ in the cSMCS model. The figures \ref{fig8} and \ref{fig9}, present a comparison of the differential cross-sections of the production of the cSMCS Higgs bosons $h_1$, $h_2$ and $h_3$ as a function of the transverse momentum at $E_{CM}=$14\;TeV, in accordance with the A1-A5 benchmarks of the cSMCS (table \ref{tab-1}). The calculations have been performed using the UPDF of KMR, illustrating the transverse momentum and rapidity distributions of these differential cross-sections. Parts (a), (b) and (c) correspond to the SM-like $h_1$ (also see figure \ref{fig4}(c)), $h_2$ and $h_3$ Higgs bosons in the cSMCS model, respectively. The uncertainty regions for these results are also included in these figures. To compute these uncertainties, we have manipulated the hard-scale $\mu$ by a factor of $2$. It is apparent that the behavior of these predictions is rather diverse, covering different kinematic regions.
{{
The differential cross-section for the $h_2$ and $h_3$ are generally smaller than the differential cross-section for $h_1$. The upper and the lower curves for $h_2$ from $A3$ and $A4$ benchmarks with a spread of $10^{-2}$ pb$/$GeV to $10^{-5}$ pb$/$GeV, respectively and for $h_3$ from $A5$ and $A1$ benchmarks with a spread of $10^{-1}$ pb$/$GeV to $10^{-3}$ pb$/$GeV, respectively.
In the figure \ref{fig9}, the general behavior of the curves is similar to the figure \ref{fig8} with a single difference: the upper curve in the $h_3$ case belongs to the benchmark $A5$. The $h_2$ and the $h_3$ curves have a spread of $10^{-5}$ pb to $10^{-7}$ pb and $10^{-3}$ pb to $10^{-5}$ pb, respectively.}}

 We believe that results presented in the figures \ref{fig8} and \ref{fig9} are reliable estimations for the Higgs boson signals within the cSMCS and will be particularly useful in the on-going experimental research regarding light and heavy Higgs bosons at the LHC.

\section{Conclusions}
\label{sec:Conclusions}
In this work, we have calculated the production rates for the SM Higgs bosons and the Higgs bosons from the cSMCS, using an effective LO partonic matrix element and the UPDF of the KMR formalism. The calculations for the SM Higgs boson have been compared with the existing experimental data of the CMS and the ATLAS collaborations and the results of a number of collinear frameworks, showing that our computations, within the given uncertainty bounds, present an acceptable platform to describe the Higgs production at the LHC. { {We have also demonstrated that the behavior of the SM-like $h_1$ Higgs boson (from any given cSMCS benchmark) is similar to the SM Higss particle.}} Afterward, we have presented our predictions regarding the distribution of the transverse momentum and the rapidity of the produced SM-like $h_1$ and heavy Higgs bosons $h_2$ and $h_3$, from the cSMCS at the LHC. Detecting heavier Higgs bosons, if happen, will open the doors for further exploration of these ideas. These predictions may provide some insight regarding the dynamics of the next discovery. { {If a newly discovered scalar boson happens to follow one of these kinematic behaviors, one might potentially have a clue of the behavior of the next discovery (hence confirming the validity of the cSMCS in describing the physics beyond the SM). }}

\begin{acknowledgments}

The authors sincerely Professor M. Krawczyk for reviewing this work and for instructive comments. 

ND is supported in part by the National Science Center, Poland, the HARMONIA project under contract UMO-2015/18/M/ST2/00518.

MRM acknowledges R. Aminzadeh-Nik for his valuable discussions and comments.

\end{acknowledgments}

\appendix 
\section{Matrix Element}
\label{AppB}
Deriving an analytic expression for ${\mathcal{M}}$ is mathematically involved, specially with non-negligible transverse momenta for the incoming gluons. Using the Feynman rules for the diagram (a) of the figure \ref{fig1} (and its permutation), one can readily write the corresponding matrix element as:
\begin{eqnarray}
	\mathcal{M}(g^{*}(k_1)&&+g^{*}(k_2) \to H(p))  = 2\pi \alpha_S(\mu^2) \int_0^{\infty} {d^4q \over (2\pi)^4}
	\nonumber \\ && \times
	m_t \left( { G_F \over \sqrt{2}} \right)^{1/2}  Tr \left\{ {\gamma.(q+k_2) + m_t \over (q+k_2)^2 - m_t^2} \right.
	\nonumber \\ && \times 
	 \left[ \gamma_{\mu} \epsilon_a^{\mu} (\lambda_1,\textbf{k}_1) t^a \right]
	 {\gamma.(q) + m_t \over q^2 - m_t^2}
	 \left[ \gamma_{\nu} \epsilon_b^{\nu} (\lambda_2,\textbf{k}_2) t^b \right]
	 \nonumber \\ && \times
	 \left. {\gamma.(q-k_1) + m_t \over (q-k_1)^2 - m_t^2} + [k_1 \leftrightarrow k_2]\right\} .
	 \label{eq9}
\end{eqnarray}
In the equation (\ref{eq9}), $\alpha_S$ and $G_F$ are respectively the running coupling constant of the strong interaction and the Fermi's constant. $q$ is the 4-momenta of the exchanged particle in the top-quark loop. $\epsilon_a^{\mu} (\lambda_i,\textbf{k}_i)$ are the polarization functions of the incoming gluons, relative to their spin state ($\lambda_i$), their momenta and their color state (denoted by the color index $a$) and $t^a$, the Gell-Mann matrices which are the generators of the $SU(3)$ gauge group. To calculate $|\mathcal{M}|^2$, one have to multiply the expression (\ref{eq9}) by its complex conjugate, do the traces and perform the integration. Additionally, since the incoming gluons are virtual, one have to take into account the so-called non-sense polarization (see the references \cite{NLO-W/Z,Deak}) through the following identity:
\begin{equation}
	\sum_{\lambda} \epsilon^{\mu} (\lambda, \textbf{k}_i) \epsilon^{* \nu} (\lambda, \textbf{k}_i) =
	{ k_{i,t}^{\mu} k_{i,t}^{\nu} \over \textbf{k}_{i,t}^2}.
	\label{eq10}
\end{equation}
Hence, after rather lengthy calculations, one obtains:
\begin{eqnarray}
	|\mathcal{M}|^2 && = {\alpha_S^2(\mu^2) \over 288\pi^2} {G_F \over \sqrt{2}}
	 \; \tau^2 \; |D(\tau)|^2 \; ( m_H^2 + {p}_t^2 )^2 \; cos^2 \varphi ,
	 \nonumber \\
	\label{eq11-2}
\end{eqnarray}
where $\tau = 4m_t^2/m_H^2$ and
\begin{eqnarray}
	D(\tau < 1) && = {3 \over 2} \left[ 1 + {1 - \tau \over 2 } 
	\left( ln {1 + \sqrt{1-\tau} \over 1 - \sqrt{1-\tau}} - i\pi \right)^2 \right], 
	\nonumber \\ 
	D(\tau \geq 1) && = {3 \over 2} \left[ 1 + (1 - \tau) arcsin^2 
	\left( {1 \over \sqrt{\tau}} \right) \right].
	\label{eq12}
\end{eqnarray}
It is easy to confirm that at the limit of large $m_t$, i.e. as $\tau \to \infty$, 
\begin{equation}
	\lim_{\tau \to \infty} \tau D(\tau) = 1 + \mathcal{O} \left( {1 \over \tau} \right).
	\label{eq13}
\end{equation}
Afterward, neglecting the transverse momentum of the produced boson, $p_t \to 0$, the equation (\ref{eq11}) returns to its conventional from in the collinear approximation, \cite{Hautmann}.


\begin{thebibliography}{a}
\addcontentsline{toc}{chapter}{Bibliographie}

\bibitem{WattWZ} G. Watt, A.D. Martin and M.G. Ryskin, Phys.Rev.D 70 (2004) 014012.
\bibitem{Lipatov} A.V. Lipatov, N.P. Zotov, Eur. Phys. J. C 44 (2005) 559-566.
\bibitem{Luszczak:2005xs} 
  M.~Luszczak and A.~Szczurek,
  Eur.\ Phys.\ J.\ C {\bf 46}, 123 (2006)
  doi:10.1140/epjc/s2005-02464-9
  [hep-ph/0504119].
\bibitem{Grazzini} M. Grazzini, CERN-PH-TH/2004-107, arXiv:hep-ph/0406156.
\bibitem{Chen} X. Chen, J. Cruz-Martinez, T. Gehrmann, E.W.N. Glover, M. Jaquier, IPPP/16/73, ZU-TH 29/16, FR-PHENO-2016-014, arXiv:1607.08817.
\bibitem{Ferrera} G. Ferrera, J. Pires, MPP-2016-152, TIF-UNIMI-2016-7, arXiv:1609.01691.
\bibitem{Monni} P. F. Monni, E. Re, P. Torrielli, Phys. Rev. Lett. 116 (2016) 242001.
\bibitem{Florian} D. de Florian, DESY 16-107, FR-PHENO-2016-007, ICAS 08/16, MITP/16-061, ZU-TH 20/16, arXiv:1606.09519.
\bibitem{Liebler} S. Liebler, H. Mantler, M. Wiesemann, DESY 16-146, KA-TP-24-2016, ZU-TH 27/16, arXiv:1608.02949.
\bibitem{Caola} F. Caola, S. Forte, S. Marzani, C. Muselli, G. Vita, JHEP 08 (2016) 150.
\bibitem{Telnov:2013zfa} 
  V.~I.~Telnov,
  PoS IHEP{-LHC-2012}, 018 (2012)
  [arXiv:1307.3893 [physics.acc-ph]].

  \bibitem{Darvishi:2016gvm}
  N.~Darvishi and M.~Krawczyk,
  arXiv:1603.00598 [hep-ph].
 
\bibitem{Darvishi:2016tni} 
  N.~Darvishi,
  JHEP {\bf 1611}, 065 (2016)
  doi:10.1007/JHEP11(2016)065
  [arXiv:1608.02820 [hep-ph]].

\bibitem{Krawczyk:2015xhl}
  M.~Krawczyk, N.~Darvishi and D.~Sokolowska,
  Acta Phys.\ Polon.\ B {\bf 47} (2016) 183
  doi:10.5506/APhysPolB.47.183
  [arXiv:1512.06437 [hep-ph]].
 \bibitem{Bonilla:2014xba} 
  C.~Bonilla, D.~Sokolowska, N.~Darvishi, J.~L.~Diaz-Cruz and M.~Krawczyk,
  arXiv:1412.8730 [hep-ph].
\bibitem{NDthesis} N. Darvishi, Extension of the Standard Model with a Doublet and a Complex Singlet, Ph.D. Thesis, University of Warsaw, Poland, 2017. 
\bibitem{NLO-Higgs} M. Modarres, M.R. Masouminia, 2017 (in preparation).
  
\bibitem{DGLAP1} V.N. Gribov and L.N. Lipatov, Yad. Fiz., 15 (1972) 781..
\bibitem{DGLAP2} L.N. Lipatov, Sov.J.Nucl.Phys., 20 (1975) 94.
\bibitem{DGLAP3} G. Altarelli and G. Parisi, Nucl.Phys.B, 126 (1977) 298.
\bibitem{DGLAP4} Y.L. Dokshitzer, Sov.Phys.JETP, 46 (1977) 641.
\bibitem{KMR} M.A. Kimber, A.D. Martin and M.G. Ryskin, Phys.Rev.D, 63 (2001) 114027.
\bibitem{MRW} A.D. Martin, M.G. Ryskin, G. Watt, Eur.Phys.J.C, 66 (2010) 163.
\bibitem{KKMS} M.A. Kimber, J. Kwiecinski, A.D. Martin, A.M. Stasto, Phys.Rev.D 62, (2000) 094006.
\bibitem{KIMBER} M.A. Kimber, Unintegrated Parton Distributions, Ph.D. Thesis, University of Durham, U.K. (2001).
\bibitem{CCFM1} M. Ciafaloni, Nucl.Phys.B, 296 (1988) 49.
\bibitem{CCFM2} S. Catani, F. Fiorani, and G. Marchesini, Phys.Lett.B, 234 (1990) 339.
\bibitem{CCFM3} S. Catani, F. Fiorani, and G. Marchesini, Nucl.Phys.B, 336 (1990) 18.
\bibitem{CCFM4} M. G. Marchesini, Proceedings of the Workshop QCD at 200 TeV Erice, Italy, edited by L.
Cifarelli and Yu.L. Dokshitzer, Plenum, New York (1992) 183.
\bibitem{CCFM5} G. Marchesini, Nucl.Phys.B, 445 (1995) 49.
\bibitem{BFKL1} V.S. Fadin, E.A. Kuraev and L.N. Lipatov, Phys. Lett. B, 60 (1975) 50.
\bibitem{BFKL2} L.N. Lipatov, Sov.J.Nucl.Phys., 23 (1976) 642.
\bibitem{BFKL3} E.A. Kuraev, L.N. Lipatov and V.S. Fadin, Sov. Phys. JETP, 44 (1976) 45.
\bibitem{BFKL4} E.A. Kuraev, L.N. Lipatov and V.S. Fadin, Sov. Phys. JETP, 45 (1977) 199.
\bibitem{BFKL5} Ya.Ya. Balitsky and L.N. Lipatov, Sov.J.Nucl.Phys., 28 (1978) 822.
\bibitem{FL} M. Modarres, H. Hosseinkhani, N. Olanj and M.R. Masouminia, Eur. Phys. J. C 75 (2015) 556.
\bibitem{Dipole-FL} M. Modarres, M.R. Masouminia, H. Hosseinkhani, and N. Olanj, Nucl. Phys. A 945(2016)168–185.
\bibitem{NLO-W/Z} M. Modarres, M.R. Masouminia, R. Aminzadeh Nik, H. Hosseinkhani, N. Olanj, Phys. Rev. D 94 (2016) 074035.
\bibitem{W/Z-13} M. Modarres, M.R. Masouminia, R. Aminzadeh Nik, arXiv:1610.02635.
\bibitem{Dijet} M. Modarres, M.R. Masouminia, arXiv:1610.02777.
\bibitem{CMS8H} V. Khachatryan et al. [CMS Collaboration], Eur. Phys. J. C 76 (2016) 13.
\bibitem{ATLAS8H} G. Aad et al. [ATLAS Collaboration], JHEP 1409 (2014) 112.
\bibitem{ATLAS13H} ATLAS Collaboration, ATLAS-CONF-2016-067.
\bibitem{Bechtle:2008jh} 
  P.~Bechtle, O.~Brein, S.~Heinemeyer, G.~Weiglein and K.~E.~Williams,
  Comput.\ Phys.\ Commun.\  {\bf 181}, 138 (2010)
  doi:10.1016/j.cpc.2009.09.003
  [arXiv:0811.4169 [hep-ph]].
\bibitem{ref1}
  G.~Aad {\it et al.} [ATLAS Collaboration],
  Phys.\ Rev.\ D {\bf 90} (2014) no.11,  112015
  doi:10.1103/PhysRevD.90.112015
  [arXiv:1408.7084 [hep-ex]].

\bibitem{ref2}
  V.~Khachatryan {\it et al.} [CMS Collaboration],
  Eur.\ Phys.\ J.\ C {\bf 74} (2014) no.10,  3076
  doi:10.1140/epjc/s10052-014-3076-z
  [arXiv:1407.0558 [hep-ex]].
  
\bibitem{ref3}
  M.~Baak {\it et al.}  [Gfitter Group Collaboration],
  Eur.\ Phys.\ J.\ C {\bf 74} (2014) 9,  3046
  [arXiv:1407.3792 [hep-ph]].
\bibitem{kt-fact} S. Catani, M. Ciafoloni, F. Hautmann, Nucl. Phys. B 366, 135 (1991).
\bibitem{AOC} M.A. Kimber, A.D. Martin and M.G. Ryskin, Eur. Phys. J. C12 (2000) 655.
\bibitem{PNLO} W. Furmanski, R. Petronzio, Phys. Lett. B 97, 437 (1980).
\bibitem{VEGAS}G. P. Lepage, J. Comput. Phys. 27 (1978) 192.
\bibitem{MMHT2014} L. A. Harland-Lang, A. D. Martin, P. Motylinski, R.S. Thorne, Eur.Phys.J.C 75 (2015) 204.
\bibitem{Deak} M. Deak, Transversal momentum of the electroweak gauge boson and forward jets in high energy factorisation at the LHC, Ph.D. Thesis, University of Hamburg, Germany, 2009.
\bibitem{Hautmann} F. Hautmann, Phys. Lett. B535, 159 (2002).

\end{thebibliography}
\end{document}